\documentclass[9pt,twocolumn,twoside]{osajnl}
\pdfoutput=1
\journal{ol} 

\setboolean{shortarticle}{true} 

\title{Improvements of computational ghost imaging by using Special-Hadamard patterns}

\author[1,2,]{Jie Hou}
\author[1]{Yuan Sun}
\author[1]{Wanqing Yang}
\author[1]{Ting Lv}
\author[1,*]{Xiaoqian Wang}

\affil[1]{Department of Physics, Changchun University of Science and Technology, Changchun 130022, People’s Republic of China}
\affil[2]{Department of Physics and Engineering, Delaware State University, Dover, Delaware 19901, USA}

\affil[*]{Corresponding author: xqwang21@163.com}

\ociscodes{(110.1758) Computational imaging; (110.2970) Image
detection systems; (100.2980) Image enhancement.}

\begin{abstract}
We introduced a new kind of patterns named Special-Hadamard patterns, which could be used as structured illuminations of computational ghost imaging. Special-Hadamard patterns can  get a better image quality than Hadamard patterns in a noisy environment. We can completely reconstruct the original object in a noiseless environment by using Special-Hadamard patterns, and the size of object also can be adjusted arbitrarily,  these advantages cannot be achieved by other common patterns. We also performed simulations to compare the results of Special Hadamard patterns with the results of Hadamard patterns. We found Special Hadamard patterns can greatly improve the image quality of computational ghost imaging.
\end{abstract}

\setboolean{displaycopyright}{true}

\begin{document}

\maketitle

Computational ghost imaging (CGI) proposed by Shapiro is based on correlation measurements between the structured illumination and the total intensity transmitted (or reflected) by an object in 2008\cite{erkmen2010ghost,shapiro2012physics,
shapiro2008computational,bromberg2009ghost}. CGI has many  advantages and attracts a lot of attention in various applications\cite{Clemente:10,chen2013object,sun20133d,welsh2013fast}.
In recent years, many algorithms of CGI have been proposed to improve its computing efficiency and imaging quality\cite{sun2012normalized,xu20181000,yao2014iterative}, such as differential ghost imaging\cite{ferri2010differential}, compressed sensing ghost imaging\cite{katz2009compressive,katkovnik2012compressive,zerom2011entangled}, and so on. The pre-programmed patterns also significantly affect the computing efficiency and the imaging quality, and recently more and more researchers begin to focus on them\cite{wang2016influence,khamoushi2015sinusoidal,yang2016scalar,zhang2015ghost,
gao2017optimization}.

In CGI, researchers usually use a projector to generate structured patterns, and it is found that random binary patterns  can get excellent imaging quality in experiment. Hadamard pattern is a special binary pattern. Hadamard pattern computational ghost  imaging (HCGI) has significant advantages over  random binary patterns in CGI system. It can be reconstructed from  the original image almost completely without algorithm noise. HCGI also has strong robustness with less number of measurements. Hadamard pattern is used extensively in experiments and lots of papers related to Hadamard patterns has been published recently
\cite{tetsuno2017subpixel,wang2016fast}. 

However, HCGI has some drawbacks which restricting image quality and application range. In this paper, we investigate the causes of these drawbacks. And then we introduce a new kind of patterns Special-Hadamard patterns to solve these problems. We evaluate the result of Special-Hadamard matrix computational ghost imaging (SHCGI). We also compared the numerical simulations of HCGI and SHCGI. The reconstracted image quality is improved by using Special-Hadamard patterns, and these patterns are more universal than Hadamard patterns.

Hadamard matrix (H) is a square matrix whose entries are either $+1$ or $-1$ and whose rows and columns are mutually orthogonal independently. Then it has some special properties for an $N$-dimensional Hadamard matrix, $HH^{\mathrm{T}}=NI$ ($I$ is the  identity matrix). $N$ is strictly limited by Hadamard matrix, \begin{equation}N\in\mathbb{A}, \mathbb{A}=\{ 2^i,12\times 2^i,20\times2^i\},\end{equation} Here, $i$ is a positive integer. We generate Hadamard matrix by computer directly in daily experiment.
In CGI system, we denote Hadamard matrix as 
\begin{equation}\label{eq:22}
H=[H_1,H_2,\dots,H_N].
\end{equation} And we project  the $n$th known pattern which is gained from reshaping  $H_n$.

For simplification, we use a vector-matrix notation with $N_1\times N_2$ images given as vectors in $T$ with dimension $N$ ($N=N_1N_2$),  we denote $T$ as \begin{equation}\label{eq:21}
T=[t_1, t_2,  \dots, t_N]^{\mathrm{T}}.
\end{equation}

\begin{figure*}  
\centering  
\includegraphics[width=\linewidth]{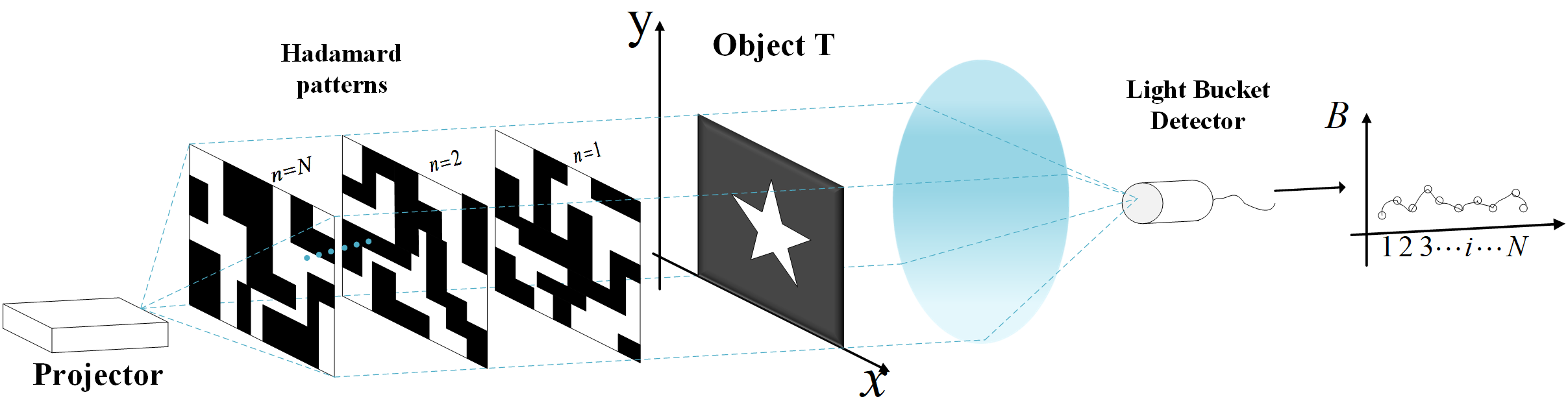}  
\caption{Schematic  experimental setup of a typical CGI system with Hadamard patterns.}  
\label{fig:picture001}  
\end{figure*}

Fig.\ref{fig:picture001} presents the schematic of a typical CGI system with Hadamard patterns.  

The second-order correlation function of HCGI can be written as
\begin{equation}\label{eq:1}
G=\frac1N\sum_{n=1}^NH_nB_n
,\end{equation}
where $B_n$ is  the light intensity signal corresponding the $n$th pattern from the bucket sensor, and $B_n$ can be written as
\begin{equation}\label{eq:2}
B_n=H^{\mathrm{T}}_nT.
\end{equation}
According to Eqs.  (\ref{eq:1}) and (\ref{eq:2}), the second-order correlation function can be written as
\begin{equation} \label{eq:3}
G=\frac1NHH^{\mathrm{T}}T.
\end{equation}
From the definition of Hadamard matrix, we can get $HH^{\mathrm{T}}=NI$, where $I$ is the identity matrix, and then
\begin{equation}\label{eq:23}
G=\frac 1N NIT=T,
\end{equation}
that means the original image can be completely reconstructed.  

Because  the item ``$-1$'' cannot be illuminated by the projector in experiment,  we need to transform  ``$-1$'' into ``$0$'' of Hadamard matrix by the function
\begin{equation}\label{eq:24}
\widetilde H=\frac{H+1}{2}.
\end{equation}
Here, we call $\widetilde H$ pseudo-Hadamard matrix. 
For HCGI, we use the $\widetilde H$ matrix to reconstruct the imaging, and we find that
\begin{equation}\label{eq:6}
\frac 1N \widetilde H\widetilde H^{\mathrm{T}}
=\renewcommand\arraystretch{1.5}
\begin{bmatrix}
1&0.5&0.5 &\dots&0.5\\
0.5&0.5& 0.25&\dots&0.25\\
0.5& 0.25&0.5&\dots&0.25\\
\vdots&\vdots&\vdots&\ddots&\vdots\\
0.5&0.25&0.25&\dots&0.5
\end{bmatrix}_{N\times N},
\end{equation}
According to Eq. (\ref{eq:3}), for an ideal noiseless environment, the reconstructed image $G$ of HCGI can be written as
\begin{equation}\label{f:TGI}
G=\frac 14 (\sum_{n=1}^{N}t_{n}+t_1)+\frac 14 
\renewcommand\arraystretch{1.2}\begin{bmatrix}
t_1+\sum\limits_{n=1}^{N}t_{n}\\
t_2\\
t_3\\
\vdots\\
t_N
\end{bmatrix}_{N\times 1}.
\end{equation}

If we want to reconstruct the original imaging completely, a relationship needs to be satisfied between $G$ and $T$, 
\begin{equation}\label{eq:manzudetiaojian}
G=\alpha T+\beta.  \qquad   (\alpha\;,\beta\; are\; constants.)
\end{equation}
According to the Eq. (\ref{f:TGI}), we can find the first pixel is bigger than others, and $G$ cannot satisfy this relationship of Eq. (\ref{eq:manzudetiaojian}). Thus we cannot get the same reconstructed image as the original. 
The reconstructed result $G$ is manully revised, removing the first error pixel to reconstruct the rest of this image. And for the remaining pixels, Eq. (\ref{eq:manzudetiaojian}) can be partly satisfied.

\begin{figure}  
\centering 
\includegraphics[width=\linewidth]{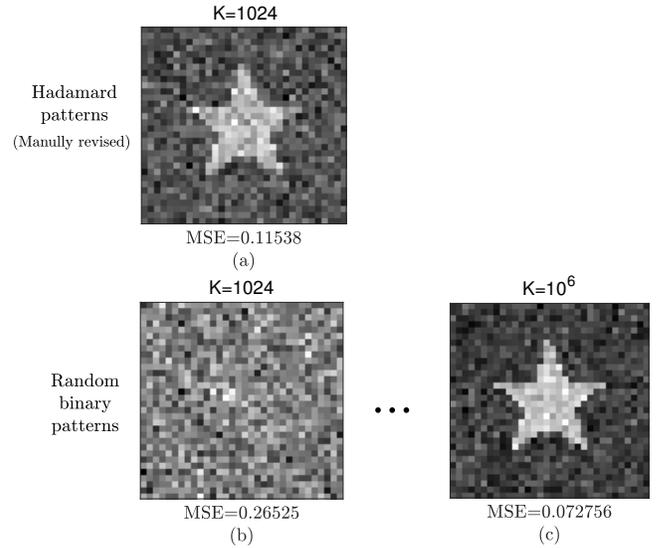}  
\caption{ Comparison of CGI results with various patterns and measurement times with the same noise level.
1024 Hadamard patterns (a), 1024 (b) and $10^6$ (c) random binary patterns.}  
\label{fig:picture002}  
\end{figure}

If we consider the system without the first pixel of HCGI, we give the constructed image of pseudo-Hadamard patterns and random binary patterns of CGI in Fig. \ref{fig:picture002},    when the system has noise. Here, we use mean square error (MSE) to descibe the image quality,
\begin{equation}
MSE=\frac 1N\sum_{n=1}^N(g(n)-T(n))^2,
\end{equation}
where \begin{equation}
g(n)=\frac{G(n)-G_{min}}{G_{max}-G_{min}},
\end{equation}
here $G_{max}$  is the maximal value of all the $G$ (except the first pixel), and $G_{min}$ is the minimal value of all the $G$. 
We find that the number of measurements are the same, the constructed image of pseudo-Hadamard patterns of CGI is better than random binary pattern of CGI. But, the number of measurements is fixed. When the number of measurements is increasing, the reconstructed image quality from random binary pattern  is better than pseudo-Hadamard pattern, which means the number of measurements is very important.

\begin{figure}  
\centering  
\includegraphics[width=\linewidth]{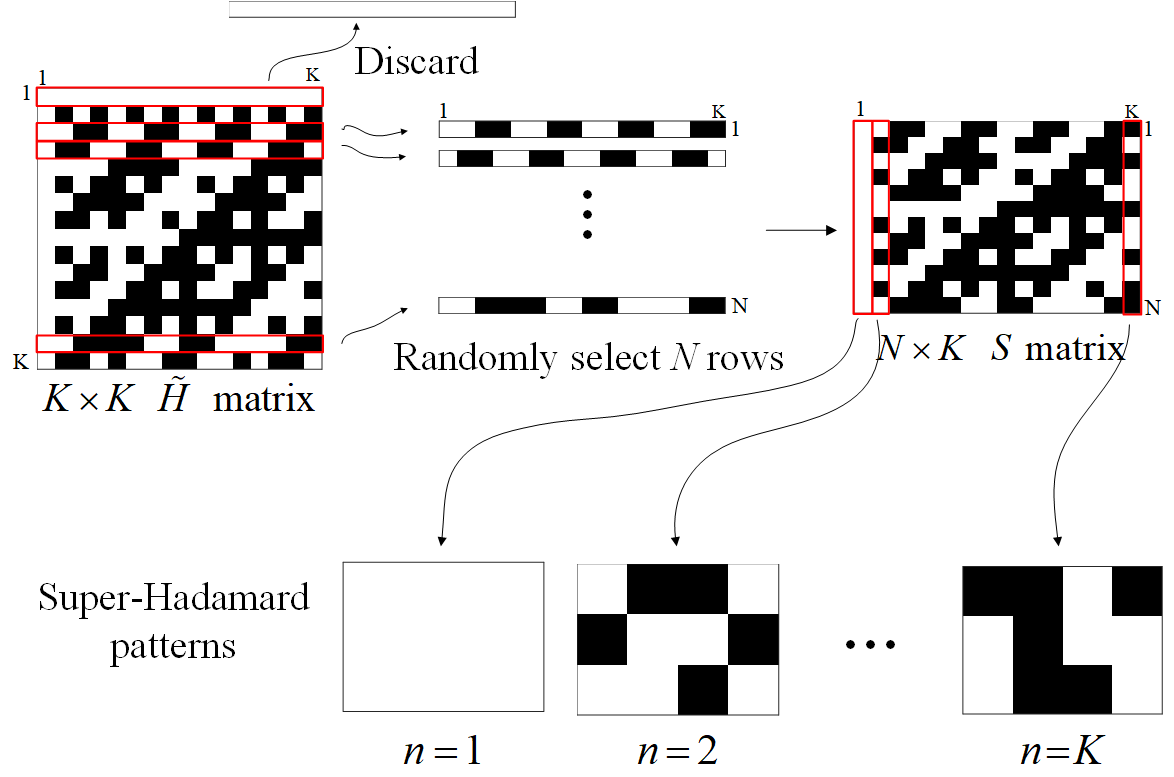}  
\caption{Procedure of generating  Special Hadamard patterns.}  
\label{fig:picture003}  
\end{figure}

We introduce a new kind of patterns Special-Hadamard patterns.
Fig. \ref{fig:picture003} show the procedure of generating Special-Hadamard patterns. 
We first generate a  $K\times K$ pseudo-Hadamard matrix ($\widetilde H$) and remove the first row of $\widetilde H$, then random extract $N$ rows from the remaining    matrix to make up an $N\times K$ Special Hadamard matrix ($S$). 
Finally, we reshape each column into projected pattern.
We also need  to ensure that the number of independent columns of  Special Hadamard matrix is more than $N$, otherwise incomplete image will be produced because of  insufficient information.

We can easily derive the characteristic of $S$ from the characteristic of $\widetilde H$.
 For an $N\times K$ Special Hadamard matrix $S$, we can get \begin{equation}\label{eq:6}
\frac 1K  SS^{\mathrm{T}}
=
\begin{bmatrix}
0.5&0.25&0.25 &\dots&0.25\\
0.25&0.5& 0.25&\dots&0.25\\
0.25& 0.25&0.5&\dots&0.25\\
\vdots&\vdots&\vdots&\ddots&\vdots\\
0.25&0.25&0.25&\dots&0.5
\end{bmatrix}_{N\times N}.
\end{equation}
Then, in a noiseless environment, the reconstructed image $G$ of SHCGI can be written as 
\begin{equation}\label{f:TGI1}
G=\frac 14 (\sum_{n=1}^{N}t_{n})+\frac 14 
\renewcommand\arraystretch{1.2}\begin{bmatrix}
t_1\\
t_2\\
t_3\\
\vdots\\
t_N
\end{bmatrix}_{N\times 1},
\end{equation}
that means the image can be reconstructed completely according to  Eq. (\ref{eq:manzudetiaojian}).

\begin{figure}
\centering  
\includegraphics[width=\linewidth]{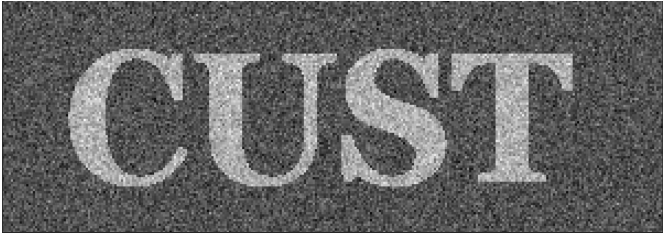}  
\caption{Reconstructed from a $100\times 300$ pixels image by SHCGI with 32768 Special-Hadamard patterns.}  
\label{fig:6}  
\end{figure}

The column length of Special Hadamard matrix ($N$) is arbitrary ($N<K$) for SHCGI, that means we can reconstruct from the original image that the number of pixels can be arbitrary. And for HCGI, the number of pixels is in set $\mathbb{A}$. As shown in Fig.  \ref{fig:6}, we reconstruct from the image which has 30000 binary transimitive pixels with 32768 Special Hadamard patterns.

\begin{figure}  
\centering  
\includegraphics[width=9cm]{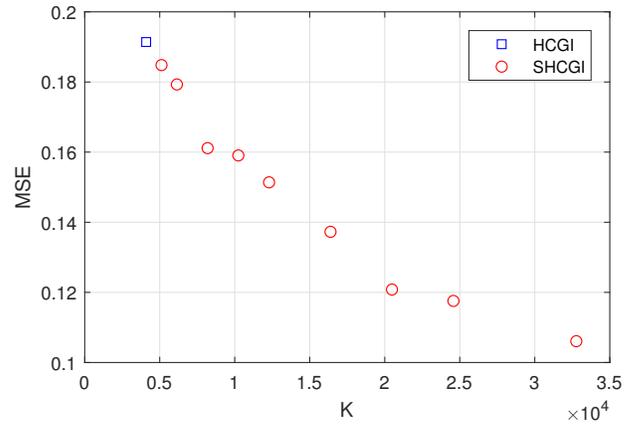}  
\caption{ MSE of the SHCGI result with different number of measurements $K$ with same  noise level.}  
\label{fig:5}  
\end{figure}  
We have discussed that the number of measurements is very significant to the quality of reconstructed image. SHCGI break the  limitation of measurement times of HCGI, then we can reconstruct from the original object with far more patterns. The number of Special Hadamard patterns ($K$) can be increased to infinity and is in set $\mathbb{A}$.
We perform numerical simuliations to study the relationship between MSE and the number of measurements. In this experiment, the original object has 4096 binary transimitive pixels, thus HCGI only has 4096 patterns.
It is found that the the quality of reconstructed images are greatly improved, as shown in Fig. \ref{fig:5}, when the number of measurements is increased.

\begin{figure*}[htbp]
\centering  
\includegraphics[width=\linewidth]{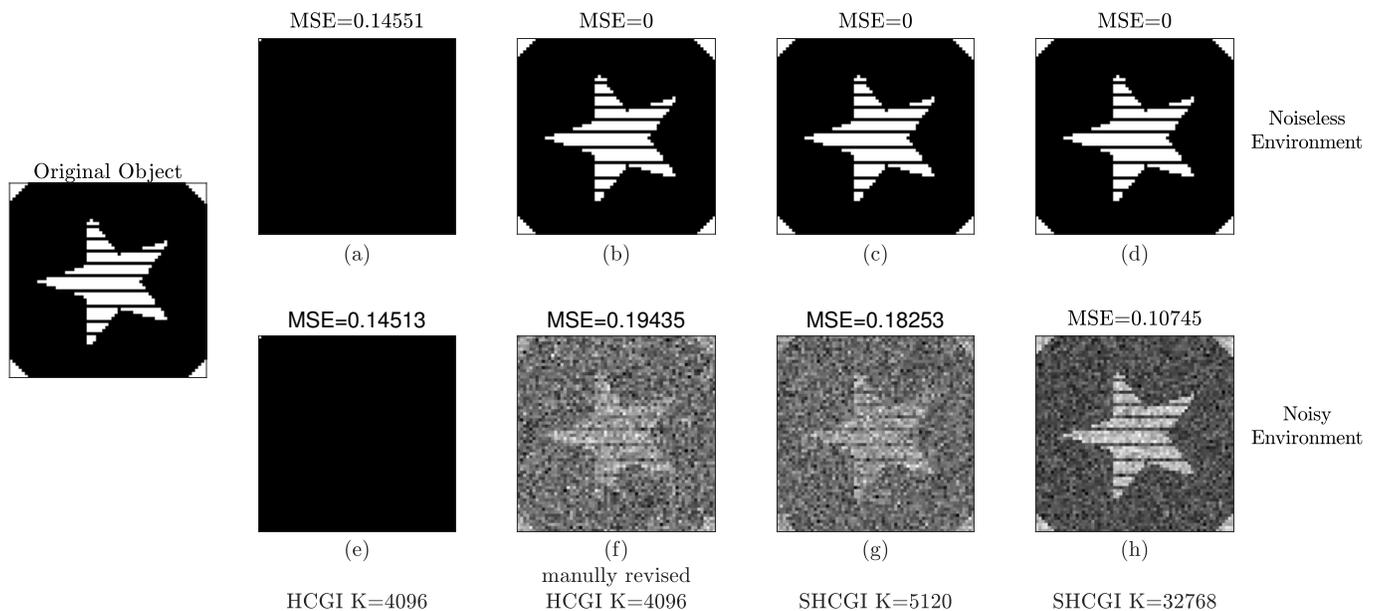}  
\caption{Comparisons of HCGI results and SHCGI results in noiseless [(a)-(d)] and noisy environments [(e)-(h)].  Results of HCGI [(a) and (e)] and results of HCGI after omitting the first pixel [(b) and (f)] with 4096 Hadamard patterns. Results of SHCGI with 5120 Special-Hadamard patterns [(c) and (g)] and 32768 patterns [(d) and (h)].}  
\label{fig:picture004}  
\end{figure*}

We also perform numerical simulations to compare SHCGI results with HCGI results, as shown in Fig. \ref{fig:picture004}. 
The original object has 4096 binary transimitive pixels.  
In a noiseless environment, we can find HCGI cannot reconstruct from the original image in Fig. \ref{fig:picture004} (a), and HCGI can restore successfully after manully revising the first pixel  in Fig. \ref{fig:picture004} (b). And SHCGI can reconstruct image completely from $MSE=0$ in Fig. \ref{fig:picture004} (c) and (d). 
In a noisy environment, the results are reconstructed from 4096 patterns, 5120 patterns and  32768 patterns, as shown in Fig. \ref{fig:picture004} (f)-(h) respectively. It is very clear that the image quality are improved when patterns are increased, that means SHCGI has more powerful robustness than HCGI.

To sum it all up, we analysed the drawbacks of restricting the reconstructed image quality in Hadamard pattern computational ghost imaging (HCGI). Then we introduced a new kind of patterns Special-Hadamard patterns, which are used as the structured illumination of computational ghost imaging. Special Hadamard patterns are generated from Hadamard matrix. But Special Hadamard pattern computational ghost imaging (SHCGI) break the HCGI limitations on the number of measurements ($K$) and the number of pixels ($N$). In SHCGI system, $K$ can be increased to infinity ($K\in \mathbb{A}$) and the total number of pixels ($N$) is arbitrary integer($N\leq K-1$), and that means SHCGI has more powerful robustness and wider range of applications than HCGI. SHCGI also can avoid the abnormal point which is founded at the first pixel of HCGI result, that means SHCGI can reconstruct the original image completely in a noiseless environment. 
In simulation, we compared SHCGI results with HCGI results using mean square error, showed that SHCGI can greatly improve the image quality of computational ghost imaging.  We suppose that these improvements will be helpful in
practical applications.


This work was supported by the National Natural Science Foundation of China (Grant No. 11305020), and the Science and Technology Research Projects of the Education Department of Jilin Province, China (Grant No. 2016-354), as well as the Science \& Technology Development Project, Jilin Province (Grand No. 20180520165JH).

\bibliography{OSA-journal-template}

\bibliographyfullrefs{OSA-journal-template}

\end{document}